\documentclass{article}
\usepackage{graphics}
\begin{document}
\title{Bandgap Extremization: Some Exact Results}
\author{Prabasaj Paul\footnote{Current address: Department of Physics and Astronomy, Colgate University, 13 Oak Drive, Hamilton NY 13346} and Bill Sutherland\\Department of Physics\\115S 1400E Room 201\\University of Utah\\Salt Lake City UT 84112}
\maketitle
\begin{abstract}
We present here a variational method for maximizing the bandgap in a 
one-dimensional system where the potential is subject to given constraints. 
Two specific examples are studied in detail. In the first, we show that if 
the potential  is constrained to lie between two values, the largest bandgap 
is obtained by a mixture of the highest and lowest potential - an exact 
result valid in any dimension. The second example fixes the first and 
second moments of the potential and seeks to extremize the bandgap. An 
exact result is obtained. Finally, we indicate how our techniques may be 
applied to photonic bandgaps.
\end{abstract}
\section{Introduction}
Over the past decade, the design and study of photonic bandgap materials 
have attracted a lot of attention \cite{website}. The design process often 
involves extremization of a bandgap with respect to variations in the 
dielectric constant profile in the material, subject to certain constraints. 
Materials so designed are expected to find use in optical communication. 
The theoretical work in this field lies in two broad areas: exact results 
that prove existence and provide bounds \cite{exact} and numerical 
algorithms that provide optimal designs \cite{numerical}. In this paper 
we present a variational argument that provides criteria that the optimal 
design must satisfy, illustrated by two examples. (The calculations presented 
here are for quantum-mechanical bandgaps; however, the results are applicable 
also to the scalar Helmholtz equation - relevant for photonic band structure 
calculations - as indicated later.) The first example leads us to a 
maximum contrast rule: if the potential is allowed to vary over a range of 
values, the optimal configuration is a combination of the largest and 
smallest values. The second example starts with a constraint on the second 
moment of the potential energy and leads to a sufficient (but not necessary) 
relationship between the wavefunction and the potential that - rather 
surprisingly - are satisfied by the solution we obtain. We verify the 
criteria and examine their consequences in one-dimension. Our exact 
results may be used as benchmarks in testing numerical algorithms and 
as tests for sharpness of proposed bounds. 

We begin with the Schr\"{o}dinger equation \cite{helm}
\begin{equation}
\label{schrod}
-\nabla^2\psi+v\psi=\epsilon\psi.
\end{equation}
Upon imposing appropriate boundary conditions on the wavefunction $\psi$, 
this selects the energy eigenvalues $\epsilon_n$ and eigenfunctions $\psi_n$. 
We assume $v$ to be real, and thus we can choose $\psi_n$ to be real. 
The index $n=0,1,2,...$ orders the energy levels, so 
$\epsilon_{n+1}\ge\epsilon_n$. In one dimension with $\psi$ real, 
this index $n$ can be taken to be the number of nodes in the wavefunction, 
or equivalently the winding number in the $\psi-\psi^\prime$ phase plane. 
In this paper - especially when we perform explicit calculations - we shall henceforth assume one dimension unless otherwise stated.

We now wish to consider the following sort of question: What potential 
$v$ - subject to constraints - gives a maximum $\epsilon_n$ for a 
particular state $n$? More elaborate questions will be considered shortly.

What type of constraints on the potential are we interested in? Since 
adding a constant $v(\vec{r})\rightarrow v(\vec{r})+v_0$ gives the trivial 
extension $\epsilon_n\rightarrow\epsilon_n+v_0$, the first constraint on 
$v$ should probably be to fix the average of $v$, so
\begin{equation}
\langle v\rangle=\int d\tau\, v(\vec{r})=v_1.
\end{equation}
The next constraint should limit the overall deviations from the 
average, and two reasonable choices might be to fix:
\begin{enumerate}
\item the {\em second moment}, so
\begin{equation}
\langle v^2\rangle=\int d\tau\, v^2(\vec{r})=v_2.
\end{equation}
\item the {\em maximum contrast}, so
\begin{equation}
v_{max}\ge v(\vec{r})\ge v_{min}.
\end{equation}
\end{enumerate}
In general, we assume a set of $J$ constraints of the form
\begin{equation}
G_j[v]=\int d\tau\, g_j[v]=v_j,\quad j=1,2,...,J.
\end{equation}
(The maximum contrast constraint can be cast into this form by an 
appropriate limiting $g[v]$; however, as we shall see, there are 
more direct methods.)

Now suppose we have solved the Schr\"{o}dinger equation (\ref{schrod}). 
We now make a small perturbation of the form
\begin{equation}
v(\vec{r})\rightarrow v(\vec{r})+\delta v(\vec{r}).
\end{equation}
We evaluate the effect of this perturbation on the energy eigenvalue, 
to first order:
\begin{equation}
\label{delE}
\delta\epsilon_n\approx\langle\delta v\rangle_n.
\end{equation}
If we seek $\epsilon_n$ to be an extremum then we require our original state to satisfy
\begin{equation}
\label{vardelv}
0=\langle\delta v\rangle_n=\int d\tau\,\delta v\psi_n^2
\end{equation}
for any allowed variation $\delta v$. As an example of constraints, let us fix the first and second moments of $v$; 
then we require
\begin{eqnarray}
0&=&\int d\tau\,\delta v\\
0&=&\int d\tau\, v\delta v.
\end{eqnarray}
Thus, (\ref{vardelv}) can be satisfied in general if
\begin{equation}
\psi_n^2=\alpha+\beta v,
\end{equation}
or
\begin{equation}
v=a+b\psi_n^2.
\end{equation}
This allows us to close the equations by solving the non-linear 
Schr\"{o}dinger equation
\begin{equation}
\label{nls}
-\nabla^2\psi_n+[a+b\psi_n^2]\psi_n=\epsilon_n\psi_n.
\end{equation}

The solution of the Schr\"{o}dinger problem with the maximum contrast 
constraint is even easier, since we see from (\ref{vardelv}) that to 
maximize $\epsilon_n$ we must choose
\begin{equation}
v=\left\{\begin{array}{l} v_{max},\mbox{ where } \psi_n^2>\phi^2\\
v_{min},\mbox{ where } \psi_n^2<\phi^2.\end{array}\right.
\end{equation}
The constant $\phi$ is chosen by imposing the first moment constraint.

In general, then, $\langle\delta v\rangle_n=0$ is satisfied by choosing
\begin{equation}
\psi_n^2=\sum_{j=1}^J\alpha_j\frac{dg_j[v]}{dv}.
\end{equation}
This equation can be inverted for $v$, giving $v=h[\psi_n^2|\alpha]$, 
where $\alpha$ represents the set of $J$ parameters $\alpha_j$. We then 
solve appropriate non-linear equations:
\begin{equation}
-\nabla^2\psi_n+h[\psi_n^2|\alpha]\psi_n=\epsilon_n\psi_n.
\end{equation}

In this paper, we wish to maximize not the energy eigenvalues of the 
Schr\"{o}dinger equation itself, but rather the energy bandgap between 
two levels, subject to constraints on the potential similar to previously. 
However, the constraints on the average of $v$ is not really necessary, 
since we consider only bandgaps. (Note that the indices on $\psi$ no 
longer refer to their ordering. In fact, we usually require that $\psi_2$, 
$\psi_1$ refer to consecutive eigenstates $\psi_{n+1}$, $\psi_n$, and so 
for periodic boundary conditions in one dimension, these wavefunctions - 
the {\em band edges} - will have exactly the same number of nodes.) Thus, 
requiring the bandgap to be an extremum with respect to constrained 
variations of $v$, we now have the following three coupled equations:
\begin{eqnarray}
-\nabla^2\psi_2+v\psi_2&=&\epsilon_2\psi_2\\
-\nabla^2\psi_1+v\psi_1&=&\epsilon_1\psi_1\\
\psi_2^2-\psi_1^2&=&\sum_{j=1}^J\alpha_j\frac{dg_j[v]}{dv}.
\end{eqnarray}
Again, we can invert to find $v$, and rewrite the problem as two coupled non-linear Schr\"{o}dinger equations
\begin{eqnarray}
-\nabla^2\psi_2+h[\psi_2^2-\psi_1^2|\alpha]\psi_2&=&\epsilon_2\psi_2\\
-\nabla^2\psi_1+h[\psi_2^2-\psi_1^2|\alpha]\psi_1&=&\epsilon_1\psi_1.
\end{eqnarray}
These equations are no longer quite so simple. We now consider our previous examples, in one dimension.

\section{Two examples}
\subsection{Maximum contrast constraint}
\begin{figure}
\includegraphics{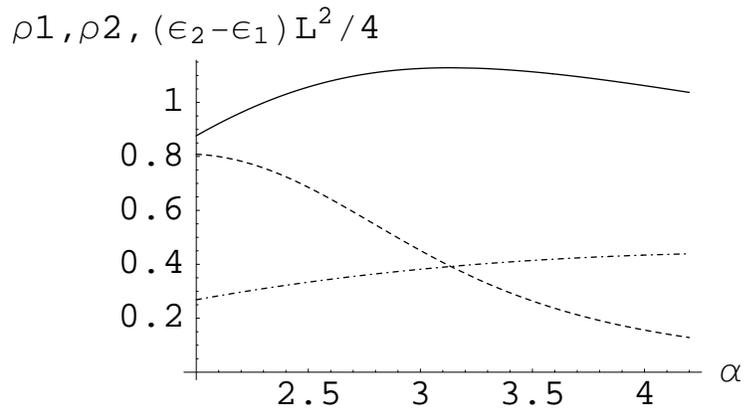}
\caption{\label{fig1}$\rho_1(A)$, $\rho_2(A)$ and $(\epsilon_2-\epsilon_1)L^2/4$ 
(dot-dash, dash and solid lines, respectively) against $\alpha$ at $\eta=5$.}
\end{figure}

\begin{figure}
\includegraphics{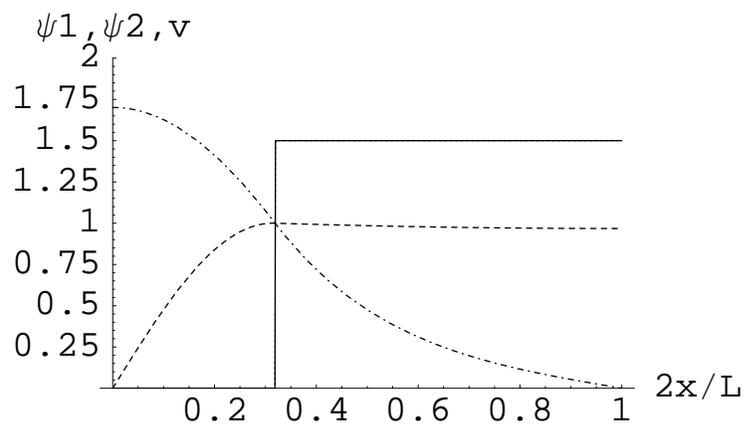}
\caption{\label{fig2}$\psi_1$, $\psi_2$ and $v$ ($v_0=1.5$) (dot-dash, dash and solid 
lines, respectively) against $2x/L$.}
\end{figure}

\begin{figure}
\includegraphics{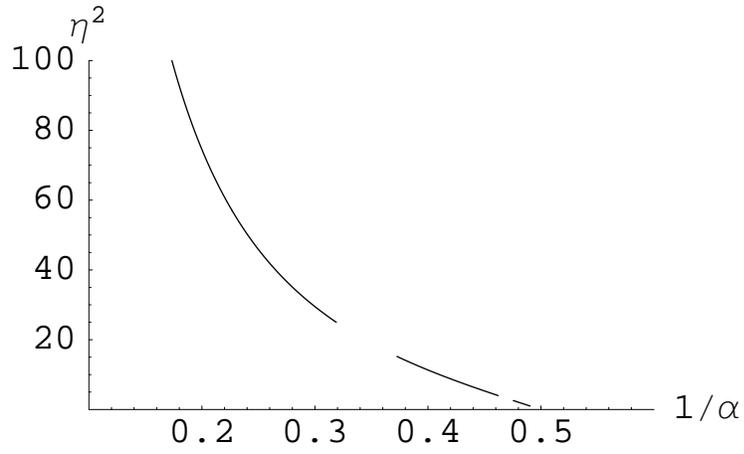}
\caption{\label{fig3}$\eta^2=v_0L^2/4$ against $1/\alpha=2A/L$.}
\end{figure}

\begin{figure}
\includegraphics{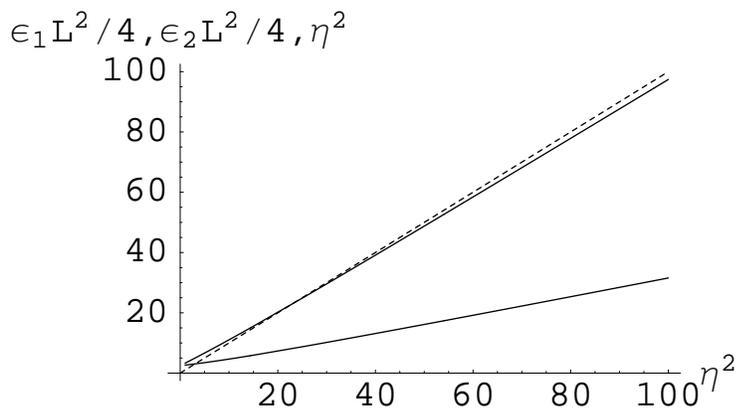}
\caption{\label{fig4}$\epsilon_1L^2/4$, $\epsilon_2L^2/4$ (lower and upper solid lines, 
respectively) and $v_0L^2/4$ against $v_0L^2/4$.}
\end{figure}

\begin{figure}
\includegraphics{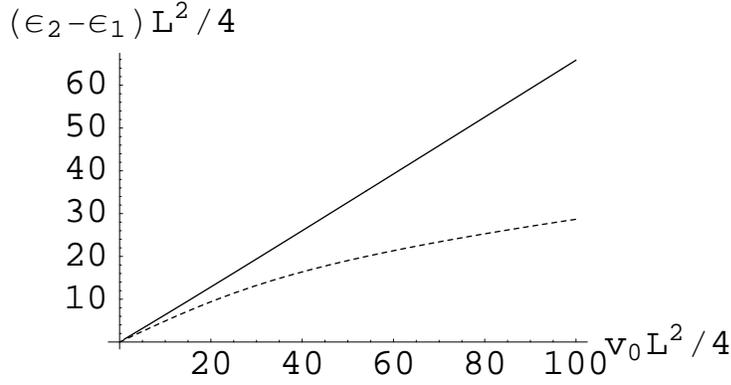}
\caption{\label{fig5}Size of bandgap (times $L^2/4$) against $v_0L^2/4$ for the optimized 
periodic square-well (solid line) and a sinusoidal potential with the same 
period and contrast (dashed line).}
\end{figure}

Without loss of generality, we reformulate the maximum contrast constraint as $v_0\ge v(x)\ge 0$. Then returning directly to (\ref{delE}), we find
\begin{equation}
\delta(\epsilon_2-\epsilon_1)=\int dx\,[\psi_2^2-\psi_1^2]\delta v\le 0
\end{equation}
with no constraint on the average of $v$. Clearly the solution is 
\begin{equation}
\label{vandrho}
v(x)=\left\{ \begin{array}{r@{,\,\mbox{if}\,}l}
v_0 & \psi_2^2(x)>\psi_1^2(x)\\ 0 & \psi_2^2(x)<\psi_1^2(x).\end{array}\right.
\end{equation}  
Furthermore, we fix $v(x)$
to be periodic with period $L$, and designate $\psi_1$ and
$\psi_2$ as each having one node per period (of $v(x)$). In
other words, the two selected wavefunctions are at the band edges
with Bloch wave vector $k=\frac{\pi}{L}$. The sole adjustable parameter
left in the potential is the width of the `barrier' of height $v_0$,
which we call $2A$ ($0\leq A\leq L/2$). Thus, we have:
$$v(x)=\left\{ \begin{array}{r@{\quad:\quad}l}
0 & -A<x<A \\ v_0 & A<x<L-A \end{array} \right.$$ extended periodically.
Symmetry dictates that $\psi_1$ has nodes at $x=mL$ and $\psi_2$,
at $x=(n+\frac{1}{2})L$ (or vice versa) where $m,n$ are integers.
Therefore,
$$\psi_1(x)=\left\{ \begin{array}{r@{\quad:\quad}l}
\cos(k_1x) & -A<x<A \\ a_1\sinh[\kappa_1(x-\frac{L}{2})] & A<x<L-A
\end{array}\right.$$ and
$$\psi_2(x)=\left\{ \begin{array}{r@{\quad:\quad}l}
\sin(k_2x) & -A<x<A \\ a_2\cosh[\kappa_2(x-\frac{L}{2})] & A<x<L-A
\end{array}\right.$$ both extended periodically. Here $k^2+\kappa^2=v_0$
and $a_i$ are constants that ensure matching at the boundaries and
may be eliminated to give:
\begin{equation}
\label{eq2}
y_1\tan y_1=\sqrt {{\frac {{\eta}^{2}}{{\alpha}^{2}}}-{y_1}^{
2}}\coth\left(\left (\alpha-1\right )\sqrt {{\frac {{\eta}^{2}}{{
\alpha}^{2}}}-{y_1}^{2}}\right)
\end{equation}
and
\begin{equation}
\label{eq1}
{\frac {\tan y_2}{y_2}}=-{\frac {1}{\sqrt {{\frac {{\eta}^{2
}}{{\alpha}^{2}}}-{y_2}^{2}}}}\coth\left(\left (\alpha-1\right )\sqrt 
{{\frac {{
\eta}^{2}}{{\alpha}^{2}}}-{y_2}^{2}}\right)
\end{equation}
where $y_i\equiv k_iA$, $\alpha\equiv L/2A$ and $\eta\equiv\sqrt{v_0}L/2$.
The variables to be solved for are $y_1,y_2$ and $\alpha$. A third equation 
is provided by (\ref{vandrho}) which implies that $\psi_1^2(x)=\psi_2^2(x)$ 
at $x=A$ (and $x=L-A$):
\begin{eqnarray}
\label{eq3}
&&\cot^2 y_1\left (1-\frac{\left (\alpha-1\right ){y_1}^{2}}
{{\frac {{\eta}^{2}}{{\alpha}^{2}}}-{y_1}^{2}}\right)
-\frac{\cot y_1}{y_1}\left (1+\frac{{y_1}^{2}}{\left ({\frac {{\eta}^{2}}{{\alpha}^{2}}}-{y_1}^
{2}\right )}\right )\nonumber\\
&=&\tan^2 y_2\left (1-\frac{\left (\alpha-1\right ){y_2}^{2}}
{{\frac {{\eta}^{2}}{{\alpha}^{2}}}-{y_2}^{2}}\right)
+\frac{\tan y_2}{y_2}\left (1+\frac{{y_2}^{2}}{\left ({\frac {{\eta}^{2}}{{\alpha}^{2}}}-{y_2}^
{2}\right )}\right ). 
\end{eqnarray}
The solutions to these equations are displayed in the figures. Fig.\ref{fig1}
is a simultaneous plot of $\rho_1(A)$, $\rho_2(A)$ and 
$(\epsilon_2-\epsilon_1)L^2/4$ against $\alpha$ at
$\eta=5$. Consistent with expectation, the energy difference is an 
extremum (maximum, in this case) when $\rho_1(A)=\rho_2(A)$, at 
$\alpha=3.136$; Fig.\ref{fig2} is a simultaneous plot of $\psi_1$, 
$\psi_2$ and $v(x)$ for this set of parameter values. Fig.\ref{fig3} shows 
$\eta^2(=v_0L^2/4)$ against $1/\alpha(=2A/L)$. The breaks in the plot 
are a consequence of poor convergence of the numerical routine used 
as $\kappa_2$ and then $\kappa_1$ switch from imaginary to real with 
increasing $\eta$. Fig.\ref{fig4} shows $\epsilon_1L^2/4$, $\epsilon_2L^2/4$ 
and (for reference) $v_0L^2/4$ against $v_0L^2/4$. Finally, Fig.\ref{fig5} 
shows the energy difference (times $L^2/4$) in this case as well as for the 
potential $v(x)=v_0\cos^2\frac{\pi x}{L}$ for the same Bloch states, 
for comparison. (The latter has analytical eigenfunctions in terms of 
Mathieu functions.) As might be expected, the difference is larger with 
our optimized choice of potential.
\newcommand{\cn}{\mbox{cn}}
\subsection{Second moment constraint}
The equations we are to solve are
\begin{eqnarray}
\label{set}
-\psi_2^{\prime\prime}+v\psi_2&=&\epsilon_2\psi_2\\
-\psi_1^{\prime\prime}+v\psi_1&=&\epsilon_1\psi_1\\
\psi_2^2-\psi_1^2=\alpha v.
\end{eqnarray}
We proceed with a roundabout strategy to solve these coupled equations by 
first solving the pair of coupled equations:
\begin{eqnarray}
-\psi^{\prime\prime}+v\psi&=&\epsilon\psi\\
\psi^2&=&\beta v.
\end{eqnarray}
This is equivalent to the single equation
\begin{equation}
\label{nls1d}
\psi^{\prime\prime}-\psi^3/\beta+\epsilon\psi=0,
\end{equation}
which is (\ref{nls}) is one dimension. Suppose we find two such solutions 
with the same $v$, yet independent. Then we also have a solution to 
(\ref{set}), with $\alpha=\beta_2-\beta_1$. Clearly this gives a solution, 
but is it possible? The answer, as we demonstrate explicitly below, is yes.

\begin{figure}
\includegraphics{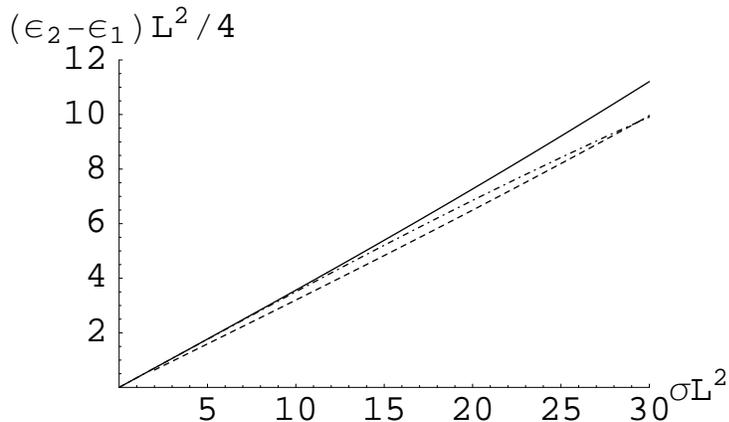}
\caption{\label{fig6}Size of bandgap (times $l^2/4$) against $\sigma L^2/4$ for the 
optimized potential (solid line), a sinusoidal potential with the same 
period and second moment (dot-dashed line) and the periodic square-well 
optimized (to a different criterion) in the previous example (dashed line).}
\end{figure}

We anticipate a periodic potential (with period $L$) and further impose the 
condition that the two energy eigenstates we seek are at the band edges, with 
period $2L$ for both. (The eigenfunctions may, then, be chosen to be real.) 
The solutions to (\ref{nls1d}) have remarkable properties 
\cite{bill,shastry}, one of which is exploited here. The equation may be 
solved in terms of the (periodic) Jacobian elliptic functions $\mbox{sn}$ 
or $\mbox{cn}$. (The third possibility, dn, turns out not to be appropriate.) 
We may now specifically require that the period of the potential be $L$. 
We finally obtain:
\begin{eqnarray}
\psi_2(x) & = &\frac{k}{\sqrt{L(1-\mbox{E}/\mbox{K})}}\mbox{sn}
(\frac{2\mbox{K}x}{L})\nonumber\\
\psi_1(x) & = &\frac{k}{\sqrt{L(1-\mbox{E}/\mbox{K})}}\mbox{cn}
(\frac{2\mbox{K}x}{L})\nonumber\\
\label{delE2}\frac{(\epsilon_2-\epsilon_1)L^2}{4} & = & (k\mbox{K})^2
\end{eqnarray}
where E and K are the complete elliptic integrals of modulus $k$. Note 
that $v$ is linearly related to both $\psi_1^2$ and $\psi_2^2$. Defining 
$\sigma^2\equiv\frac{1}{L}\int_0^Ldx\,(v^2-\langle v\rangle^2)$, we have
\begin{equation}
\label{sigma}
\sigma L^2=8\mbox{K}^2\sqrt{\frac{2(1+k^2)(1-\mbox{E}/\mbox{K})-k^2-3
(1-\mbox{E}/\mbox{K})^2}{3}}.
\end{equation}
It is, perhaps, useful to think of the whole system as being parametrized 
by $k$. The suitably scaled energy difference and strength of the potential 
(as $\sigma$) may be easily obtained from equations (\ref{delE2}) and 
(\ref{sigma}), respectively. Fig.\ref{fig6} shows 
$(\epsilon_2-\epsilon_1)L^2/4$ against $\sigma L^2$. For comparison, 
the same quantities for a sinusoidal potential and the optimized potential 
in the previous example are also plotted. As expected, our optimized 
potential gives the largest energy difference.
\section{Conclusion}
We have discussed strategies for bandgap extremization subject to two very 
different, but realistic, constraints on the potential energy function. Our 
conclusions, in both cases, are illustrated by one-dimensional examples. It 
is worthwhile pointing out that the variational argument in the first
example applies explicitly in any dimension. The second relies on the 
serendipitous existence of two solutions to the equations (\ref{set}) 
in one dimension. Whether this stroke of luck can be replicated in 
higher dimensions is, to our knowledge, an open question.

We acknowledge useful discussions with Graeme Milton and David Dobson. This 
work was supported in part by a grant from the U. S. National Science 
Foundation. 


\begin{thebibliography}{}
\bibitem{helm} A unified framework that applies to both the Schr\"{o}dinger 
and Helmholtz equations starts with the general Schr\"{o}dinger equation 
with parameter $c$: $$-\nabla^2\psi+cv\psi=\epsilon\psi.$$ The energy 
eigenvalues $\epsilon_n(c)$ are a family of functions parametrized by $c$. 
With the identification $v\rightarrow\rho$, $\psi\rightarrow u$ and identical 
boundary conditions, we can solve the Helmholtz equation $$\nabla^2u+
\omega^2\rho u=0$$ by evaluating the roots of the equations 
$\epsilon_n(-\omega_n^2)=0$. The variational techniques outlined in the 
paper may be applied to these equations as well. A detailed account will 
be presented in a forthcoming paper.
\end{thebibliography}
\end{document}